\documentclass[%
 reprint,
 amsmath,amssymb,
 aps,
]{revtex4-1}

\usepackage{caption}
\usepackage{graphicx}% Include figure files
\usepackage{bm}% bold math
\usepackage{enumerate}
\usepackage{footnote}
\usepackage[usenames, dvipsnames]{color}

\begin{document}

%\preprint{APS/123-QED}
 
\title{The role of band filling in tuning the high field phases of URu$_2$Si$_2$}

\author{M. R. Wartenbe$^{1}$, K. W. Chen$^{1}$, A. Gallagher$^{1}$, N. Harrison$^{2}$, R. D. McDonald$^{2}$, G. S. Boebinger$^{1}$, R. E. Baumbach$^{1}$}

\affiliation{
 1 National High Magnetic Field Laboratory 1800 E. Paul Dirac Drive, Tallahassee, FL 32310 \\ 2 National High Magnetic Field Laboratory, Los Alamos National Laboratory,P.O. Box 1663
  Los Alamos, NM 87545\\
 }

%\author{K. W. Chen $^{1}$}%
%\author{A. Gallagher  $^{1}$}%
%\author{R. E. Baumbach  $^{1}$}%
%\author{R. C. McDonald  $^{2}$}%
%\author{G. S. Boebinger  $^{1}$}%

%\collaboration{FSU NHMFL}%\noaffiliation

\date{\today}% It is always \today, today,

\begin{abstract}

We present a detailed study of the low temperature and high magnetic field phases in the chemical substitution series URu$_2$Si$_{2-x}$P$_x$ using electrical transport and magnetization in pulsed magnetic fields up to 65T.  Within the hidden order $x$-regime (0 \textless $\ x$ $\lesssim$ 0.035) the field induced ordering that was earlier seen for $x$ = 0 is robust, even as the hidden order temperature is suppressed.  Earlier work shows that for 0.035 $\lesssim$ $x$ $\lesssim$  0.26 there is a Kondo lattice with a no-ordered state that is replaced by antiferromagnetism for 0.26 $\lesssim$ $x$ $\lesssim$ 0.5.  We observe a simplified continuation of the field induced order in the no-order $x$-regime and an enhancement of the field induced order upon the destruction of the antiferromagnetism with magnetic field. These results closely resemble what is seen for URu$_{2-x}$Rh$_x$Si$_2$\footnote{The concentration in this paper is defined as URu$_{2-x}$Rh$_x$Si$_2$  while the chemical formula in the literature is given as U(Ru$_{1-x}$Rh$_x$)$_2$Si$_2$ [24-26]}, from which we infer that charge tuning dominantly controls the ground state of  URu$_2$Si$_2$, regardless of whether s/p or d-electrons are replaced.  Contraction of the unit cell volume may also play a role at large $x$.  This provides guidance for determining the specific factors that lead to hidden order versus magnetism in this family of materials and constrains possible models for hidden order.
\end{abstract}

\maketitle

%\tableofcontents

\section{\label{sec:level1}Introduction}
Amongst the f-electron intermetallics, URu$_{2}$Si$_2$ continues to attract interest because it hosts an unidentified ordered state (``hidden order") and unconventional superconductivity at temperatures below \textit{T$_0$} = 17.6 K and \textit{T$_c$} = 1.5 K, respectively [1-5]. These phenomena occur within a strongly hybridized f-electron lattice that is superficially similar to that of related systems with magnetically ordered ground states [6,7].  Despite this parallel, various measurements (e.g., neutron scattering) have revealed that the ordered state does not have an intrinsic magnetic moment [8]. A multitude of theories have been proposed to describe hidden order, where a distinguishing factor is the assumed degree of f-electron localization, but no consensus has been reached regarding their applicability [4,5].

 To solve this puzzle, it is important to understand what factors distinguish between the generic occurrence of magnetism in other related f-electron lattices and the singular behavior of URu$_{2}$Si$_2$. To some extent, the continuity of experimental information extracted from applied pressure (\textit{P}), chemical substitution (\textit{x}) and magnetic field (\textit{$\mu_{0}H$}) tuning series has been useful to address this question. For example, pressure drives a first order phase transition from hidden order into antiferromagnetism near \textit{P$_c$} = 5 kbar [9].  Chemical substitution also tends to promote magnetism, where Ru $\rightarrow$ Fe and Os yields phase diagrams similar to that seen with pressure [10-12], Ru $\rightarrow$ Tc and Re stabilize ferromagnetism, and Ru $\rightarrow$ Rh and Ir eventually produce antiferromagnetism [13-15].  Particularly interesting is that large magnetic fields suppress hidden order and uncover a rich family of magnetically ordered field induced (FI) states, where elastic neutron scattering in pulsed magnetic fields recently revealed that the lowest-in-magnetic field of them is a type of spin density wave order [16-20].
  
These tuning strategies reveal rich phenomena and indicate a close relationship between hidden order and magnetism, but what is missing is both a picture that unifies the diverse behavior and simple tuning schemes to access the multitude of ordered states in clean single crystals at ambient pressure. In this context, ligand site substitution in URu$_{2}$Si$_2$  is an obvious target for investigation. Thus motivated,  we recently examined the chemical substitution series URu$_2$Si$_{2-x}$P$_x$, where the Kondo lattice behavior is preserved but the hidden order (HO) is replaced by a no-ordering (NO) heavy Fermi liquid for 0.035 $\lesssim$ $x$ $\lesssim$ 0.26 that eventually gives way to antiferromagnetism (AFM) for $x$ $\gtrsim$  0.26 [21-23].  This phase diagram opens the opportunity to directly examine the effect of electronic shell filling, which at low $x$ merely tunes the density of states at the Fermi energy without disturbing the underlying band structure.

Here we report magnetoresistance and magnetization measurements in pulsed magnetic fields up to \textit{$\mu_{0}H$} = 65 T spanning the entire \textit{T-x} phase diagram of URu$_2$Si$_{2-x}$P$_x$. For concentrations in the HO $x$-regime ($x$ $\lesssim$ 0.035), the critical magnetic fields of the FI phases slightly increase, even as $T_0$ decreases.  In the NO $x$-regime (0.035 $\lesssim$ $x$ $\lesssim$ 0.26) a magnetic field induced ordered state appears for 28 T \textless \textit{$\mu_{0}H$} \textless 43 T which connects continuously to the low-$x$/large-\textit{$\mu_{0}H$} ordering. Within the AFM $x$-regime ($x$ $\gtrsim$ 0.26) magnetic fields suppress the magnetic ordering temperature towards zero, and for \textit{$\mu_{0}H$} \textgreater $\ $ 43 T an enhanced FI state appears which connects to the FI phase seen in the NO $x$-regime.  This may suggest either that the underlying Fermi surface evolves to become more favorable to magnetism at large $x$ or magnetic fluctuations are helpful to the high field ordering.

\begin{figure}
\includegraphics[scale=0.6]{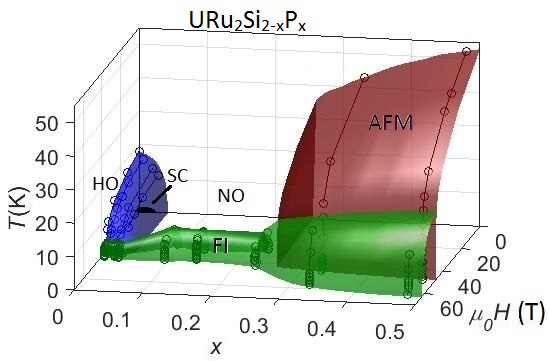}
\captionsetup{justification=raggedright,
singlelinecheck=false}
 \caption{Three dimensional phase diagram for URu$_{2}$Si$_{2-x}$P$_x$ single crystals constructed from magnetoresistance measurements, with temperature \textit{T}, magnetic field \textit{$\mu_{0}H$}, and phosphorous concentration \textit{x} as the three axes.  \textit{$\mu_{0}H$} is applied parallel to the crystallographic c-axis.  Data for $x$ and $H$ = 0 are from Ref. [21,22]. Circles are our experimental data and lines/colored regions are guides to the eye.  Regions are labeled as follows: SC = Superconductivity, HO = Hidden $x$-Order regime, NO = No-Ordered $x$-regime, FI = Field Induced order, AFM = Antiferromagnetism.}
\end{figure}
 
\begin{figure}
\includegraphics[scale=0.5]{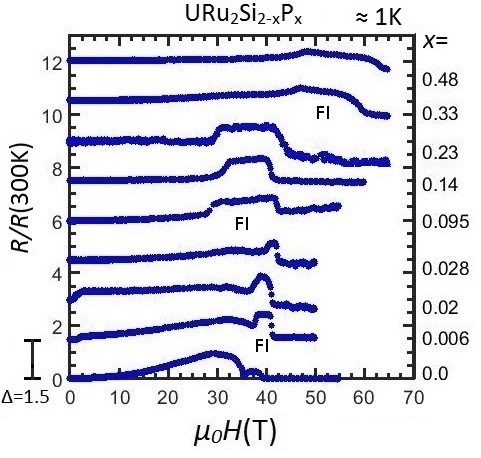}
\captionsetup{justification=raggedright,
singlelinecheck=false}
 \caption{  Waterfall plot of normalized electrical resistance \textit{R/R}(300K) vs. magnetic field \textit{$\mu_{0}H$} for various concentrations of \textit{x} at \textit{T}$\approx$1 K.  Each trace is offset by an amount $\Delta$=1.5.  The field induced FI phase originates as a narrow region in field in the HO $x$-regime (0 $\ $ \textless $\ $ $x$ $\ $ \textless $\ $ 0.035) and expands to a much broader range in the no-ordered $x$-regime (0.035 $\ $ \textless $\ $ $x$ $\ $ \textless $\ $ 0.26).  The FI order may persist at high fields in the AFM $x$-regime (0.26 $\ $ \textless $\ $ $x$ $\ $ \textless $\ $ 0.5).}
\end{figure}

\section{\label{sec:level1}Experimental}
Single crystal specimens were produced using the molten metal flux growth technique described in Refs. [21,22].  
Samples were prepared for electrical resistance measurements by spark-welding platinum wires to their surface and then gluing them to quartz substrates, after which data was collected using a 4-point AC lock-in method with magnetic field applied parallel to the c-axis.  Magnetization measurements were performed using an extraction magnetometer where mosaics of 10-20 crystals ($m$ $\approx$ 1 - 1.6 mg) were placed in Apiezon n-grease in a cylindrical plastic capsule such that their c-axis would be aligned parallel to the applied magnetic field.  Due to difficulty in loading the crystals in this configuration, they were somewhat misaligned with respect to the capsule axis and magnetic field. Measurements were made with the sample/capsule both in and out of the coil, after which the two data sets were subtracted from each other to isolate the sample/capsule signal from that of the detection coil. Electrical resistance and magnetization data were collected at temperatures 0.5K \textless $\ $ $T$ \textless $\ $ 20K and 0.6K \textless $\ $ $T$ \textless $\ $ 20K, respectively, in pulsed magnetic fields up to 65 T with pulse widths of 65 ms at the National High Magnetic Field Laboratory located at Los Alamos National Laboratory. 

\section{\label{sec:level1}Results}
Fig 1. shows the three dimensional phase diagram for single crystals of URu${_2}$Si${_{2-x}}$P${_x}$ constructed from magnetoresistance measurements, with the axes of temperature $T$, phosphorous concentration $x$, and magnetic field \textit{$\mu_{0}H$}. Data for $\textit{$\mu_{0}H$}=0$ are taken from Refs. [21,22], where the parent compound hidden order and superconductivity are rapidly suppressed for $x$ $\lesssim$ 0.035 and are replaced by a region with persistent Kondo-lattice behavior but no low temperature ordered state (NO $x$-regime). Over this $x$-range the lattice compression and strain is small and the evolution of $T_0$ and $T_c$ is attributed to s/p-shell band filling.  The NO $x$-regime persists for 0.035 $\lesssim$ $x$ $\lesssim$ 0.26, after which antiferromagnetism emerges from the f-electron lattice for 0.26 $\lesssim$ $x $ $\lesssim$ 0.5.  Starting in the middle of the NO $x$-regime, the chemical pressure $P_{ch}$ exceeds that needed to induce antiferromagnetism in the parent compound, and we infer that the influence of lattice compression becomes important over this range.  In principle, chemical disorder might also play an important role.  However, hidden order and superconductivity were previously shown to be robust even against strong disorder [22]: e.g., both ordered states persist even in specimens with residual resistivity ratios RRR = $\rho_{300K}/\rho_0$ $\approx$ 10 [32].  As shown in Supplementary Fig. 4, the residual resistivity ratio RRR $\geq$ 10 in the $x$-regime where hidden order is destroyed and is replaced by the no-order ground state [33]. From this, we infer that disorder is unlikely to be an important type of tuning.  Recent NMR measurements further elucidate the behavior in these regions, where the Kondo lattice behavior of the NO $x$-regime is similar to that seen above $T_{\rm{HO}}$ and the antiferromagnetism occurs in the bulk and has a commensurate wave vector [23].  
\par
The response of the parent compound to an applied magnetic field is also well known [16-20].  For $T$ \textless $\ $ $T_0$, the magnetoresistance initially increases with \textit{$\mu_{0}H$} and eventually drops to a minimum near 35 T, indicating the end of the hidden order phase (Fig. 2).  Within this magnetic field range, Shubnikov de Haas (SdH) oscillations reveal four regions with distinct oscillation frequencies, indicating a complex evolution of the Fermi surface [27,28].  At fields above 35T a second phase (phase II)  appears as a resistance minimum.  At approximately 36T a third phase (Phase III) appears as a step like increase in magnetoresistance, which extends up to $\approx$ 39T, before giving way to a spin polarized paramagnetic state (Phase IV).   Neutron scattering experiments in pulsed magnetic fields recently showed that phase II is an incommensurate spin density wave state with wave-vector k=(0.6,0,0)[20].  
\par
To compare with $x$ = 0, low temperature field sweeps of $\textit{R/R}$(300K) up to $\mu_{0}H$ $\lesssim$ 65T for $x$ $\lesssim$ 0.48 are shown in Fig. 2.  An important feature that is seen in these data is that some form of high field ordering persists for all $x$.  Fig. 3 details the high field ordering and summarizes resulting $T-\mu_{0}H$ phase diagrams for three concentrations spanning the $T-x$ phase diagram.  Waterfall plots of all substitutions studied can be seen in the supplementary section.  For 0 $\lesssim$ $x$ $\lesssim$ 0.035, we first see a suppression of the magnetoresistance hump with increasing $x$, which may be due to increasing charge carrier scattering due to chemical disorder.  Over this $x$-range there is a slight increase in the onset field of the FI phases and an enhancement of phase III, even as $T_0$ is suppressed (Figs. 3 a,d,g).  In the NO $x$-regime (0.035 $\lesssim$ $x$ $\lesssim$ 0.26) we unexpectedly observe a nearly-square step FI feature between 30 - 45 T (Figs. 3 b,e,h)  that appears to be a continuation of the low-$x$ field induced phases.  This feature resembles that seen in URu$_{1.92}$Rh$_{0.08}$Si$_2$, which shows zero field behavior similar to that seen in the NO $x$-regime of our series [24-26]. Within the AFM ordered regime (0.26 $\lesssim$ $x$ $\lesssim$ 0.5), an applied field suppresses the antiferromagnetism and produces a step in the magnetoresistance similar to that seen for the FI phase (Figs. 3 c,f,i).  We note that this phase is enhanced in both temperature and magnetic field range, by comparison to the lower-$x$ field induced phase.  These results might suggest that it is a distinct phase and measurements such as neutron scattering are needed to clarify this question.
 \par
 
 Fig. 4 shows waterfall plots of magnetization $M$ vs \textit{$\mu_{0}H$} for concentrations in the different regions of the phase diagram; $x$ = 0 (HO $x$-regime with FI phase), $x$ = 0.1 (NO $x$-regime with FI phase) and $x$ = 0.33 (AFM $x$-regime with FI phase).  The $x$ = 0 data, taken from Ref. [24], reveal a linear in field magnetization up to 35.8T, where a jump to 1/3 of the saturation value occurs, followed by a series of more subtle features before reaching a saturation value.  Qualitatively similar behavior is seen for $x$ = 0.1, where the data displays a 1/3 step feature at \textit{$\mu_{0}H$} = 35T.  The double step feature that is seen at 35.6T in the parent compound is absent, and the second jump to the saturation moment occurs near 40T.  This single plateau region matches with the FI phase seen in the magnetoresistance data at similar P concentration (see Fig. 2).  The $x$ = 0.33 data also reveal a  step in $M$, characterized by a broadened transition width starting around 47.5T.  After the plateau the magnetization rises again and does not reach a saturation value.  The magnetization plateau occurs on the same field range as the FI phase which appears at high field past the AFM phase in magnetoresistance measurements, and is most likely due to magnetic ordering. 
 \par
 
 \begin{figure*}
 \includegraphics[scale=0.25]{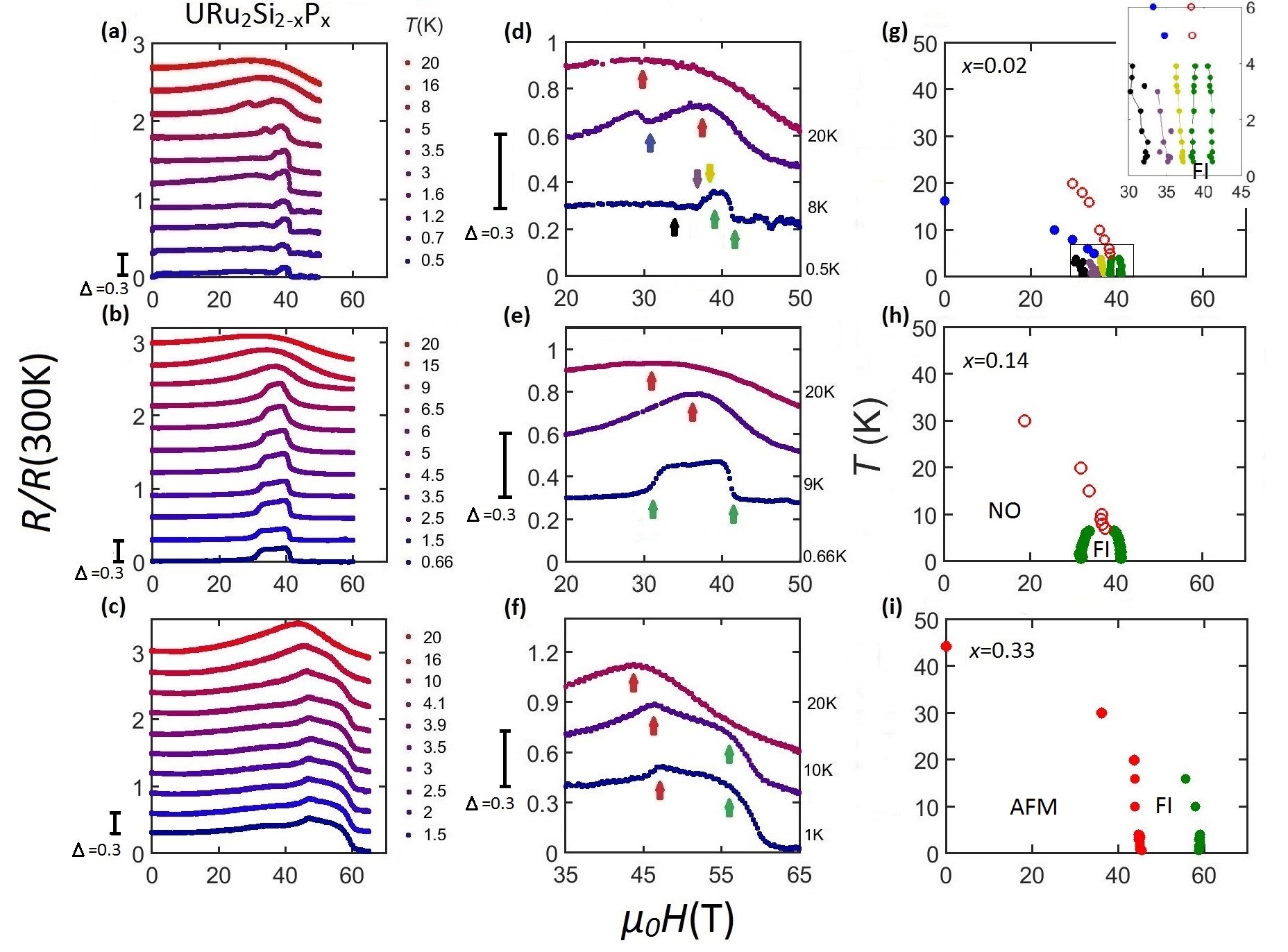}
 \captionsetup{justification=raggedright,
 singlelinecheck=false}
  \caption{  Representative data from the three regimes; hidden order $x$-regime (\textit{x}=0.02), no order $x$-regime (\textit{x}=0.14) and the antiferromagnetic $x$-regime (\textit{x}=0.33).  Normalized resistance \textit{R/R}(300K) vs. field \textit{$\mu_{0}H$} plots are shown in panels (a)-(c), the data are offset vertically by a constant amount $\Delta$ indicated in each panel.  Panels (d)-(f) highlight the FI ordering from panels in (a)-(c).  Colored arrows indicate phase transitions.   Panels (g)-(i) show the \textit{T-\textit{$\mu_{0}H$}} phase diagrams, where phase boundary are determined following the conventions from Ref. [19].  The inset region in panel (g) is the region of the cascade of phase transitions defined previously}
  \end{figure*}

  \begin{figure}
 \includegraphics[scale=0.4]{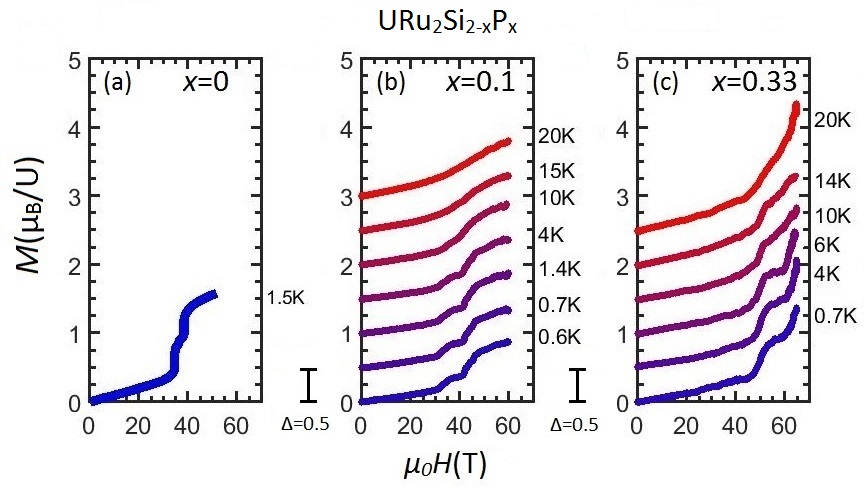}
 \captionsetup{justification=raggedright,
 singlelinecheck=false}
  \caption{ Waterfall plots of the magnetization $M$ vs magnetic field $\mu_0 H$ data. Data in panel (a) are taken from Ref. [24]. Panels (b) and (c) summarize data for two substitutions, representing materials in the no-ordered (\textit{x}=0.1) and antiferromagnetic (\textit{x}=0.33) $x$-regimes, respectively.  Data are offset by $\Delta$=0.5 for clarity.}
 \end{figure}
 
  \begin{figure}
 \includegraphics[scale=0.4]{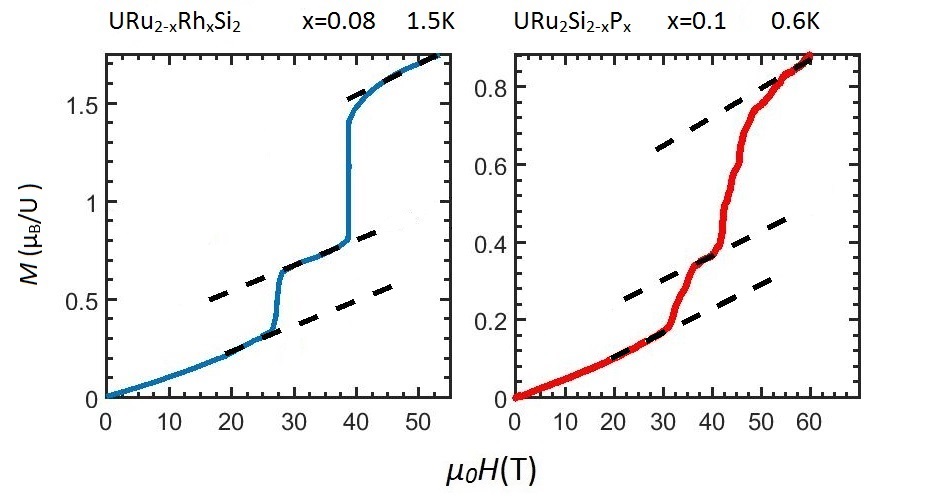}
 \captionsetup{justification=raggedright,
 singlelinecheck=false}
  \caption{  Magnetization versus magnetic field $\mu_{0}H$ for Rh-substituted [24] and P-substituted URu$_2$Si$_2$ at low temperatures in the no-order $x$-regime.  Dashed lines are guides to the eye to illustrate the 1/3 and 2/3 magnetization jumps seen in each material.}
 \end{figure}
 
 \section{\label{sec:level1}Discussion}
From these measurements we construct the $T-x-\mu_0H$ phase diagram for URu$_2$Si$_{2-x}$P$_x$, which features significant complexity with high field ordering persisting across the entire substitution series (Fig. 1). In the HO $x$-regime, the FI phases retain many of the characteristic features of the parent compound. This is even as the hidden order temperature is suppressed, suggesting that the FI behavior is not solely tied to the field driven collapse of HO.  Within the NO $x$-regime, the onset of the FI state expands to lower $\mu_0H$, but terminates at high $\mu_0H$ values similar to what is seen in the HO $x$-regime. Furthermore, the qualitative features of the magnetization field dependence are similar to what is seen in the HO $x$-regime: there are plateau regions in the magnetization with similar step sizes for both $x$ = 0 and 0.1. An attractive explanation for this is that related types of FI order emerge in the HO and NO $x$-regimes when a non-symmetry breaking electronic crossover originating from the hybridization between the f- and conduction electron states is suppressed towards zero temperature, where additional small features are seen for $x \textless$ 0.03 that relate to the suppression of HO.  Here the crossover is represented by the magnetoresistance maximum $\rho_{\rm{max}}$.  Similar behavior is seen in several other strongly correlated metals without zero field ordered ground states including CeRu$_2$Si$_2$, UPt$_3$, and Sr$_2$Ru$_3$O$_7$ [31,32]. 

Measurements that target the order parameters are needed to distinguish between the ordered states in this phase diagram. Nonetheless, some insight is gained by considering the similarities between the s/p (Si $\rightarrow$ P) and d-shell (Ru $\rightarrow$ Rh) chemical substitution series.
In Fig. 5 we compare the in-field magnetization of 4\% Rh substituted to 5\% P substituted specimens, both of which are in the NO-order $x$-region of their respective phase diagrams.  For both examples, metamagnetic jumps appear at similar fields and consist of a 1/3 jump to the first plateau and then a second 2/3 jump to the full saturation value.  Earlier work shows that the Rh substituted material orders in a ferrimagnetic up-up-down state, which is seen in Fig 5 as the first 1/3 magnetization jump [24].  We suggest that similar type of ordering may occur for the $x$ = 0.1 P substituted example.  Another intriguing feature is that the strength of the field-induced phase is enhanced by the suppression of AFM for 0.26 $\lesssim$ $x $ $\lesssim$ 0.5. The discontinuous evolution of the field and temperature extent of the FI phase indicates that magnetic fluctuations resulting from the field suppressed antiferromagnetism are involved in stabilizing this phase. Similar trends are also seen at high magnetic  fields in the antiferromagnetic regions of the $T - P$ and Ru $\rightarrow$ Fe phase diagrams, indicating a connection between these different parts of electronic phase space [29,30].

   From these measurements, we conclude that hidden order is acutely unstable against simple electronic shell filling regardless of whether it is done through chemical substitution on the d- or s/p electron sites. This is highlighted by considering that Si $\rightarrow$ P and Ru $\rightarrow$ Rh substitution might reasonably be expected to have distinct influences. For instance, they have different impact on (1) the spin orbit coupling, (2) the lattice contraction and strain, (3) the local crystal electric field, and (4) in principle could alter different parts of the Fermi surface. Furthermore, given the complexity that is seen in other chemical substitution series, under applied pressure and in high magnetic fields [9-20], a priori it seems unlikely that Si $\rightarrow$ P and Ru $\rightarrow$ Rh substitution would be equivalent. Despite this, we find phase diagrams with similar features both along the electronic shell filling and applied magnetic field tuning axes. Given that this happens on the few percent chemical substitution level and that the foundational Kondo lattice is unchanged by such small changes, it may now be possible to systematically uncover which factors underpin hidden order and thereby constrain possible theoretical models. Future measurements to probe the electronic state using advanced techniques such as angle resolved photoelectron spectroscopy and electronic Raman spectroscopy will be useful to do this.
   
This work was performed at the National High Magnetic Field Laboratory, which is supported by the U.S. National Science Foundation through NSF/DMR-1157490, the State of Florida, and the U.S. Department of Energy.  R. Baumbach, K. W. Chen, and A. Gallagher acknowledge support from the NHMFL and UCGP grant program.  N. Harrison and R. D. McDonald acknowledge support from the US Department Of Energy -BES ‘Science of 100 Tesla’.

 %\textcolor{red}{This is unexpected since there are several important energy scales in 5f-electron materials including the Kondo, RKKY, Coulomb, and spin orbit interactions and their relative importance is not obvious[33]. Even less clear is how each of them would be influenced by band filling, let alone replacement of s/p vs. d electrons. This work arranges these interactions in terms of their influence: (1).} %
%We find that the high temperature Kondo lattice behavior is insensitive to electronic tuning and while it sets the stage for low temperature interactions, it does not determine the ground state [21-23].  

 %\textcolor{red}{Through analogy to Ru $\rightarrow$ Rh series, we suggest that the high field ordered states seen in the Si $\rightarrow$ P series is a form of spin density wave magnetism.}

\section*{\label{sec:citeref}Citations and References}

\begin{enumerate}

\item T. T. M. Palstra, A. A. Menovsky, J. van den Berg, A. J. Dirkmaat, P. H. Kes, G. J. Nieuwenhuys, J. A. Mydosh, Physical Review Letters 55(24), 2727-2730 (1985).\par

\item W. Schlabitz, J. Baumann, B. Pollit, U. Rauchschwalbe, H. M. Mayer, U. Ahlheim, C. D. Bredl, Zeitschrift fur Physik B 62(2), 171-177 (1986).\par

\item M. B. Maple, J. W. Chen, Y. Dalichaouch, T. Kohara, C. Rossel, M. S. Torikachvili, M.W. McElfresh, J. D. Thompson, Physical Review Letters 56(2), 185-188 (1986).\par

\item J. A. Mydosh, P. M. Oppeneer, Reviews of Modern Physics 83(4), 01-1322 (2011).\par

\item J. A. Mydosh, P. M. Oppeneer, Philosophical Magazine 94(32-33), 3642-3662 (2014).\par

\item T. T. M. Palstra, A. A. Menovsky, G. J. Nieuwenhuys, J. A. Mydosh, Journal of Magnetism and Magnetic Materials 54-57(Part 1), 435-436 (1986).\par

\item T. Endstra, G. J. Nieuwenhuys, J. A. Mydosh, Physical Review B 48(13), 9595-9605 (1993).\par

\item P. Das, R. E. Baumbach, K. Huang, M. B. Maple, Y. Zhao, J. S. Helton, J. W. Lynn, E. D. Bauer, M. Janoschek, New Journal of Physics, Vol. 15, 053031 (2013).\par

\item M. W. McElfresh, J. D. Thompson, J. O. Willis, M. B. Maple, T. Kohara, M. S.
Torikachvili, Physical Review B 35(1), 43-47 (1987).\par

\item N. Kanchanavatee, M. Janoschek, R. E. Baumbach, J. J. Hamlin, D. A. Zocco, K. Huang, M. B. Maple, Physical Review B. 84(24), 245122 (2011).\par

\item N. Kanchanavatee, B. D. White, V. W. Burnett, M. B. Maple, Philosophical Magazine 94(32-33), 3681-3690 (2014).\par

\item P. Das, N. Kanchanavatee, J. S. Helton, K. Huang, R. E. Baumbach, E. D. Bauer, B. D.
White, V. W. Burnett, M. B. Maple, J. W. Lynn, M. Janoschek, Physical Review B. 91(8),
085122 (2015).\par

\item Y. Dalichaouch, M. B. Maple, J. W. Chen, T. Kohara, C. Rossel, M. S. Torikachvili, A. L. Giorgi, Physical Review B 41(4), 1829-1836 (1990).\par

\item N. P. Butch, M. B. Maple, Journal of Physics: Condensed Matter 22(16),
164204 (2010).\par

\item Y. Dalichaouch, M. B. Maple, M. S. Torikachvili, A. L. Giorgi, Physical Review B 39(4),
2423-2431 (1989).\par

\item M. Jaime, K. H. Kim, G. Jorge, S. McCall, J. A. Mydosh, Physical Review Letters 89(28), 287201 (2002).\par

\item J. S. Kim, D. Hall, P. Kumar, G. R. Stewart, Physical Review B. 67(1), 014404 (2003).\par

\item K. H. Kim, N. Harrison, M. Jaime, G. S. Boebinger, J. A. Mydosh, Physical Review
Letters 91(25), 256401 (2003).\par

\item G. W. Scheerer, W. Knafo, D. Aoki, G. Ballon, A. Mari, D. Vignolles, J. Flouquet, Physical Review B 85, 094402 (2012).\par

\item W. Knafo, F. Duc, F. Bourdarot, K. Kuwahara, H. Nojiri, D. Aoki, J. Billette, P. Frings, X. Tonon, E. Leli\`{e}vre-Berna, J. Flouquet, L.-P. Regnault, Nature Communications 7, 13075 doi: 10.1038/ncomms13075 (2016). \par

\item A. Gallagher, K.-W. Chen, C. M. Moir, S. K. Cary, F. Kametani, N. Kikugawa, D. Graf, T. E. Albrecht-Schmitt, S. C. Riggs, A. Shekhter, R. E. Baumbach, Nature Communications 7, 10712 (2016)
 \par

\item A. Gallagher, K.-W. Chen, S. K. Cary, F. Kametani, D. Graf, T. E. Albrecht-Schmitt, A. Shekhter, R. E. Baumbach, Journal of Physics: Condensed Matter, 29, 024004 (2017). \par

\item K. R. Shirer, M. Lawson, T. Kissikov, B. T. Bush, A. Gallagher, K.-W. Chen, R. E. Baumbach, N. J. Curro, Physical Review B 95, 041107(R) (2017). \par

\item K. Kuwahara, S. Yoshii, H. Nojiri, D. Aoki, W. Knafo, F. Duc, X. Fabr\`{e}ges, G.W. Scheerer, P. Frings, G. L. J. A. Rikken, F. Bourdarot, L. P. Regnault, J. Flouquet, Physical Review Letters 110, 216406 (2013). \par

\item K. H. Kim, N. Harrison, H. Amitsuka, G. A. Jorge, M. Jaime, J. A. Mydosh, Physical Review Letters, Vol. 93, Number 20, 206402 (2004).

\item K. H. Kim, N. Harrison, H. Amitsuka, G. A. Jorge, M. Jaime, J. A. Mydosh, arXiv:cond-mat/0411068v1 \par

\item M. M. Altarawneh, N. Harrison, S. E. Sebastian, L. Balicas, P. H. Tobash, J. D. Thompson, F. Ronning, E. D. Bauer, Physical Review Letters, Vol. 106, 146403 (2011).

\item E. Hassinger, G. Knebel, T. D. Matsuda, D. Aoki, V. Taufour, J. Flouquet, Physical Review Letters, Vol. 105, 216409 (2010).

\item Dai Aoki, Fr\'{e}d\'{e}ric Bourdarot, Elena Hassinger, Georg Knebel, Atsushi Miyake, St\'{e}phane Raymond, Valentin Taufour, Jacques Flouquet, Journal of the Physical Society of Japan, Vol. 78, 053701 (2009).

\item S. Ran, I. Jeon, N. Kanchanavatee, K. Haunf, M. B. Maple, A. Gallagher, K. Chen, D. Graf, R. Baumbach, J. Singelton, \textit{Phase diagram of URu$_{2-x}$Fe$_x$Si$_2$ under high magnetic field},  March Meeting 2017, F20.00007.\par

\item Fuminori Honda, Tetsuya Takeuchi, Shinichi Yasui, Yuki Taga, Shingo Yoshiuchi, Yusuke Hirose, Yoshiharu Tomooka, Kiyohiro Sugiyama, Masayuki Hagiwara, Koichi Kindo, Rikio Settai, Yoshichika \={O}nuki,  Journal of the Physical Society of Japan, Vol. 82, 084705 (2013).\\
\par

\item S. A. Grigera, R. S. Perry, A. J. Schofield, M. Chiao, S. R. Julian, G. G. Lonzarich, S. I. Ikeda, Y. Maeno, A. J. Mills, A. P. Mackenzie, Science, Vol. 294, (2001). \par

\item SUPPLEMENTARY

\item Kevin T. Moore, Gerrit van Laan, Reviews of Modern Physics, Vol. 81, (2009).\par

\item  R. E. Baumbach, Z. Fisk, F. Ronning, R. Movshovich, J. D. Thompson, E. D. Bauer, Philosophical Magazine, Vol. 94 (2014).

\item E. Hassinger, G. Knebel, K. Izawa, P. Lejay, B. Salce, J. Flouquet, Physical Review B, Vol. 77 (2008).

\item S. Ran, C. T. Wolowiec, I. Jeon, N. Pouse, N. Kanchanavatee, K. Huang, D. Martien, T. DaPron, D. Snow, M. Williamsen, S. Spagna, M. B. Maple, arXiv:1604.00983v1 [cond-mat.str-el].  

 \end{enumerate}

\end{document}